# Data privacy protection in microscopic image analysis for material data mining

Boyuan Ma [a, b, c, d, ¶], Xiang Yin [a, c, ¶], Xiaojuan Ban [a, b, c, d, *], Haiyou Huang [a, c, d, e, **], Neng Zhang [a, c, e], Hao Wang [a, f], Weihua Xue [f, g]

a. Beijing Advanced Innovation Center for Materials Genome Engineering, University of Science and Technology Beijing, Beijing, 100083, China.
b. Institute of Artificial Intelligence, University of Science and Technology Beijing, Beijing, 100083, China.
c. Beijing Key Laboratory of Knowledge Engineering for Materials Science, Beijing, 100083, China.
d. Shunde Graduate School, University of Science and Technology Beijing, Foshan 528300, China.
e. Institute for Advanced Materials and Technology, University of Science and Technology Beijing, Beijing, 100083, China.
f. School of Materials Science and Engineering, University of Science and Technology Beijing, Beijing, 100083, China.
g. School of Materials Science and Technology, Liaoning Technical University, Liaoning, 114051, China.

* Corresponding author at: School of Computer and Communication Engineering, University of Science and Technology Beijing, Xueyuan Road 30, Haidian District, Beijing 100083, China. Email address: banxj@ustb.edu.cn.

** Corresponding author at: Institute for Advanced Materials and Technology, University of Science and Technology Beijing, Xueyuan Road 30, Haidian District, Beijing 100083, China. Email address: huanghy@mater.ustb.edu.cn.

¶ These authors contributed equally to the work.

# Abstract

Recent progress in material data mining has been driven by high-capacity models trained on large datasets. However, collecting experimental data has been extremely costly owing to the amount of human effort and expertise required. Therefore, material researchers are often reluctant to easily disclose their private data, which leads to the problem of data island, and it is difficult to collect a large amount of data to train high-quality models. In this study, a material microstructure image feature extraction algorithm FedTransfer based on data privacy protection is proposed. The core contributions are as follows: 1) the federated learning algorithm is introduced into the polycrystalline microstructure image segmentation task to make full use of different user data to carry out machine learning, break the data island and improve the model generalization ability under the condition of ensuring the privacy and security of user data; 2) A data sharing strategy based on style transfer is proposed. By sharing style information of image that is not urgent for user confidentiality, it can reduce the performance penalty caused by the distribution difference of data among different users.

# Introduction

In the past decade, a new interdisciplinary research field called materials informatics, which combines data and information science with materials science, has led to an increasing number of successful materials discoveries[1]-[6]. With the development of machine learning, many data-driven algorithms have been proposed and widely used in many fields. As we know, for deep neural networks, the scale of training dataset and the accuracy of the trained machine learning model are usually positively correlated under the premise of ensuring data quality, the lacking of training data will make the model lack generalization ability. Now a days, the scholars who devote themselves to the research of new materials usually face the dilemma of lacking large scale of high-quality data to construct reliable machine learning models[7]. A possible way is to centralize each researcher's local data to form a large-scale public dataset. In fact, this method it is not feasible due to some privacy issues.

Federated learning is a machine learning setting where many clients collaboratively train a model under the orchestration of a central server while keeping the training data decentralized[8]. It can promote cooperation between different clients while mitigating many of the systemic privacy risks, which provide a feasible solution to break the isolated data island in new material fields. The difference between centralized training and federated training is shown in Fig 1.

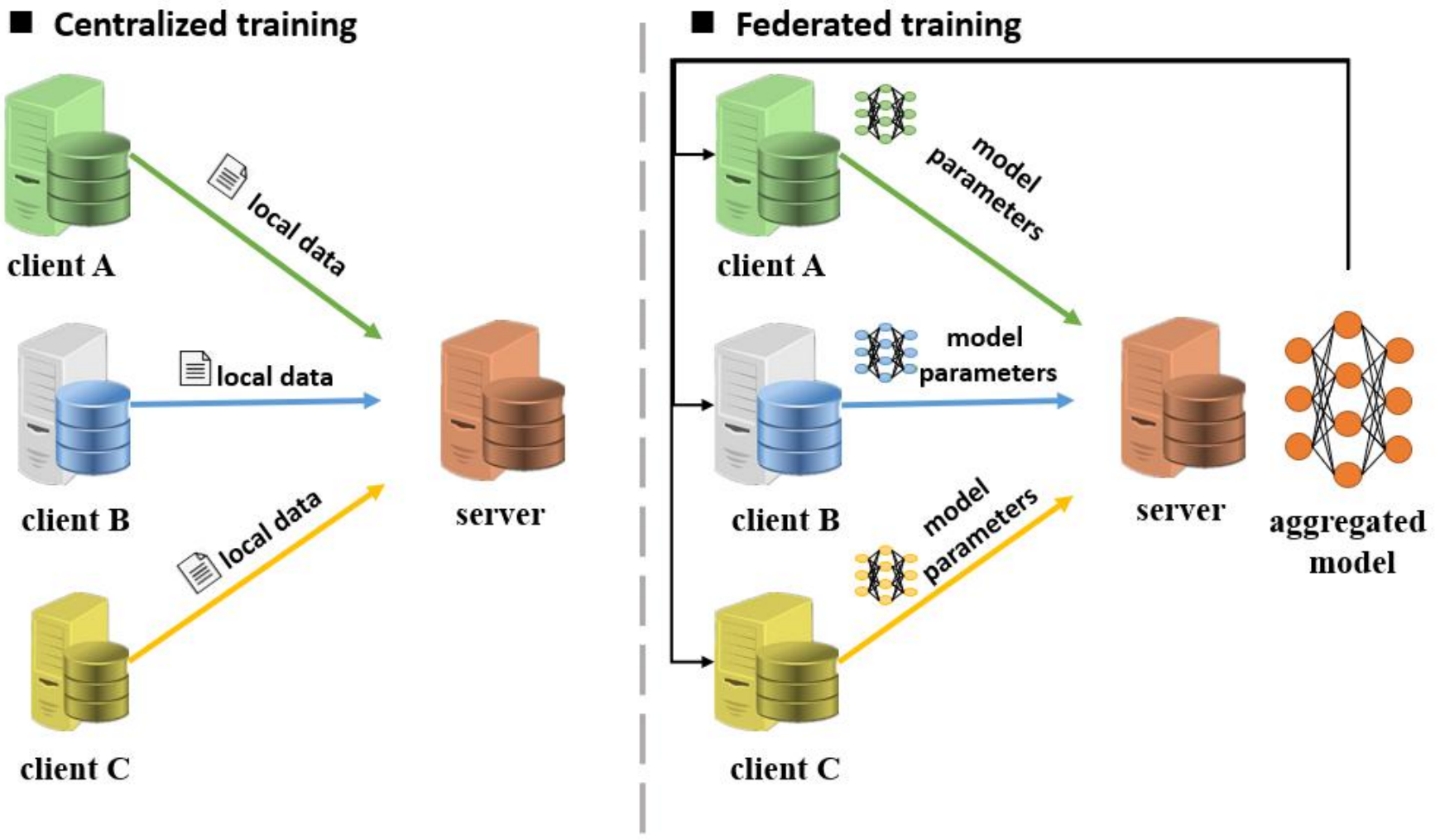

**Fig 1** The difference between centralized training and federated training

Federated learning relies on stochastic gradient descent works well when the training data between different client is independent and identically distributed(IID). But in the field of materials science, different element composition and preparation procession of material always bring difference to each researcher's local data which leads to Non-IID across different clients. Non-IID data bring challenge to federated learning[8]-[12] result in worse performance and slower convergence. Zhao proposed a data sharing strategy to address this problem[12], a global sharing dataset which consists of public dataset of specific task is distribute to every client to slightly reducing the EMD between different client's local model, experiment showed that little globally shared data brings obviously performance improvement. But this kind of data sharing method is not suitable for materials informatics researchers due to the lack of large-scale public dataset.

Material micro-structure data is an important type of material data to build the intrinsic relationship between composition, structure, process, and properties that is fundamental to material design[13]. For the researchers who want to control the material properties and performances of metals or alloys in specific industrial application, it is essential for them to perform quantitative analysis of micro-structures[14][15]. One of the key steps in the design process is microscopic image processing using computational algorithms and tools to efficiently extract useful information from the images[16], such as extract significant information from poly-crystalline iron or austenite in order to detect and depart each grain from raw microscopic images to obtain accurate descriptions of the micro-structure using image segmentation methods which generates pixel-wise labels of a raw image, the schematic diagram of deep learning based material microscopic image segmentation is shown in Fig 2.

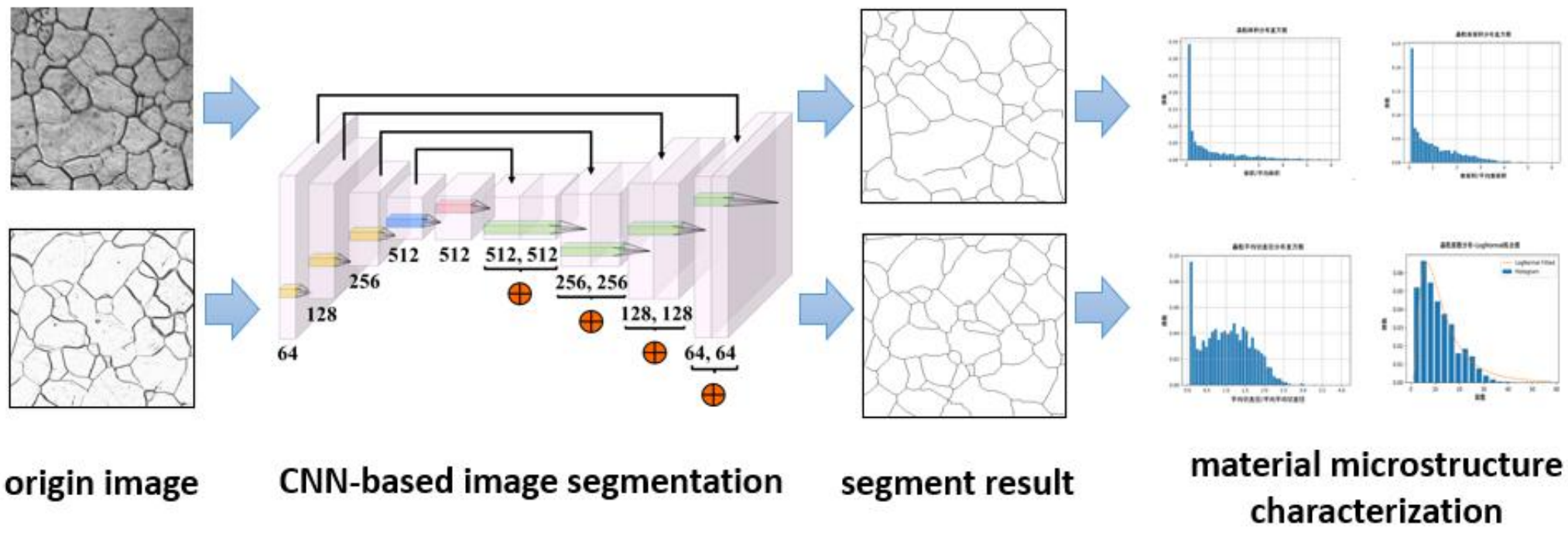

**Fig 2** Schematic diagram of deep learning-based material microscopic image segmentation

We conducted experiments on material microscopic image segmentation task. Our methods base on three assumptions: 1) material micro-structure image (such as poly-crystalline iron and austenite) data usually consist of structure information and style information as shown in Fig 3. 2)Structure information which mainly contained in the ground truth label is privacy sensitive because it contains the main information of a specific client's data on material microscopic image segmentation task. Besides, the style information is not privacy sensitive because it mainly influenced by the lighting conditions and it can be extract by training a GAN. 3) If the image is preprocessed to the same scale than the Non-IID is mainly caused by the difference in style information as shown in Fig 3. Based on the three assumptions, we develop a novel method called *FedTransfer* whose central idea is to perform data sharing by only sharing insensitive style information between different clients. Our method realizes the data sharing based on transfer learning theory. The whole method consists of two parts: federated image style transformation and federated training. The procession of federated image style transfer is shown in Fig 4 while the pseudo code is shown in algorithm 1. We used the image style transfer model pix2pixHD[17] to extract the image style information from different clients. Each client first trains a image style transfer model with its own private image dataset. In details, in order to learn the style information of the client's dataset, the generator is given a real label as input and is trained to recovery the real image, the discriminator is trained to distinguish the "real ones" versus the generated image. After that, every clients upload their image style transfer model to server and the server exchange their models and send other clients' model back. Finally, as shown in Fig 4, each client performed image-to-image transfer by converting their own segmentation labels to synthetic images with the models trained by others. After finishing federated image style transfer, clients perform federated training to collaborate with each other under the standard federated training approach (FedAvg)[18] where each client trains a copy of the global model locally on its own dataset (real data and synthetic data) and sends a weight update to the central server every communication round and server averages all updated model weights and re-deploys it as a new model to the individual client[18], this procedure is repeated until the model converges.

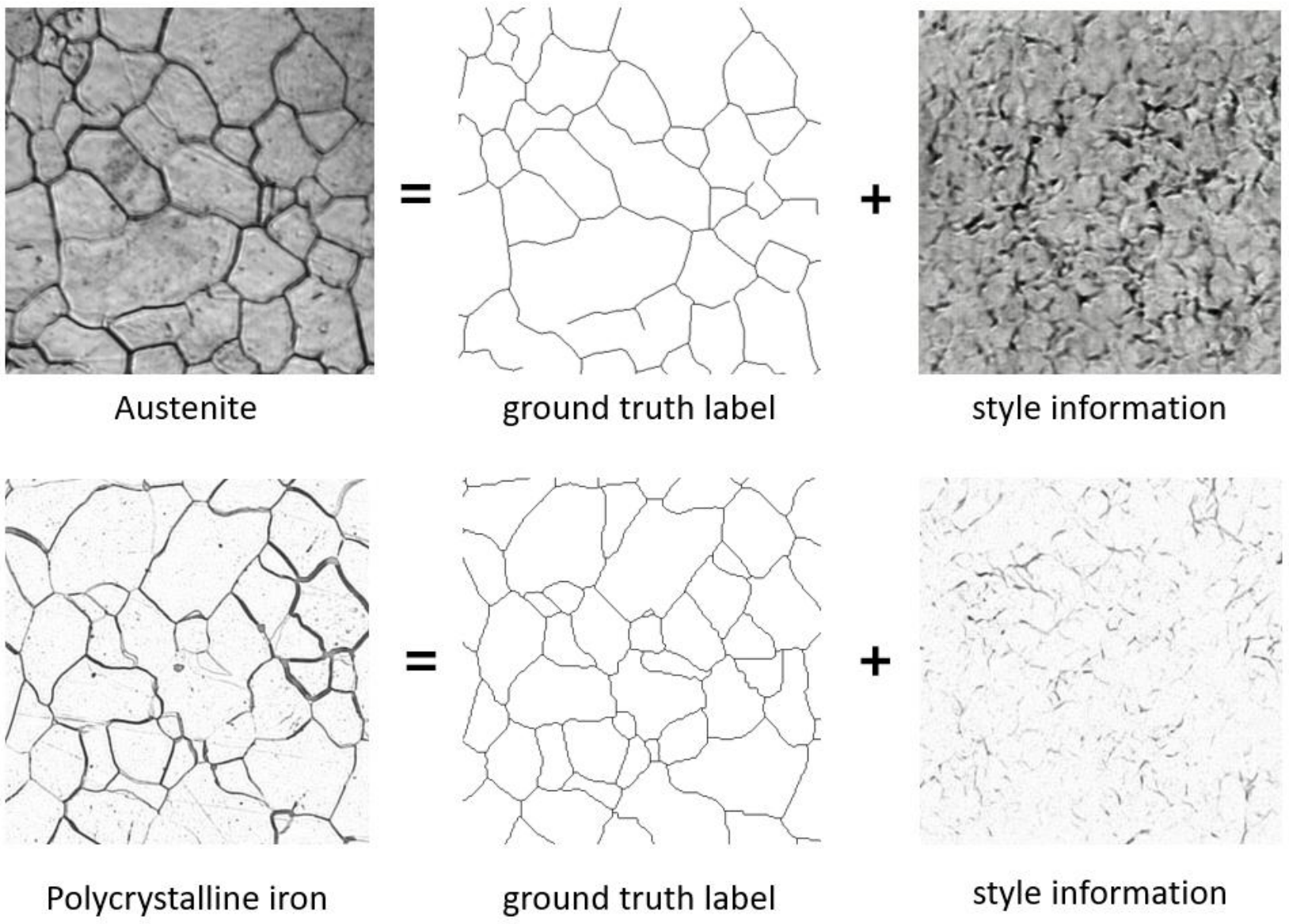

**Fig 3** The decomposition of material micro-structure image data

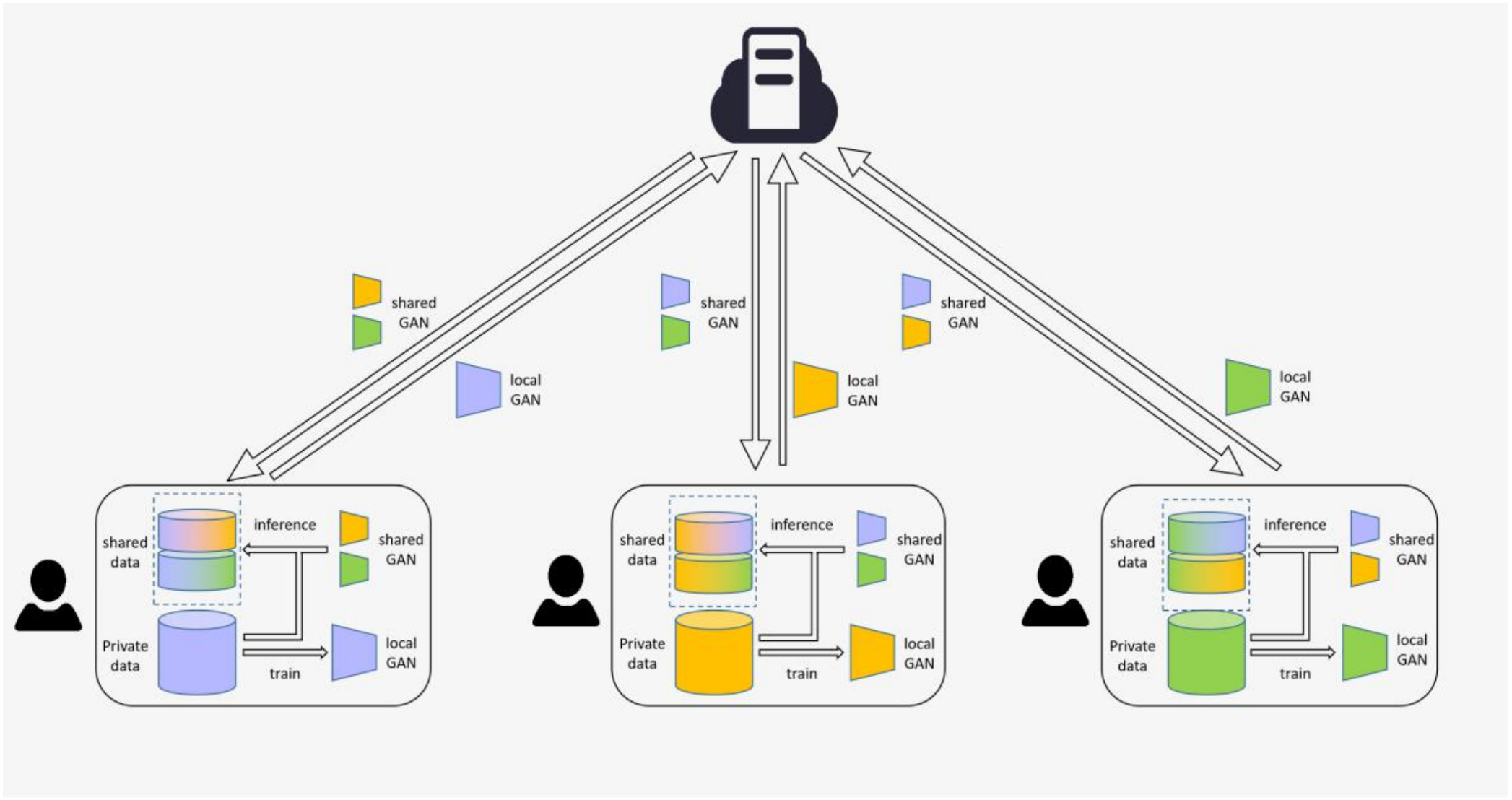

**Fig 4** The flow chart of federated image style transfer

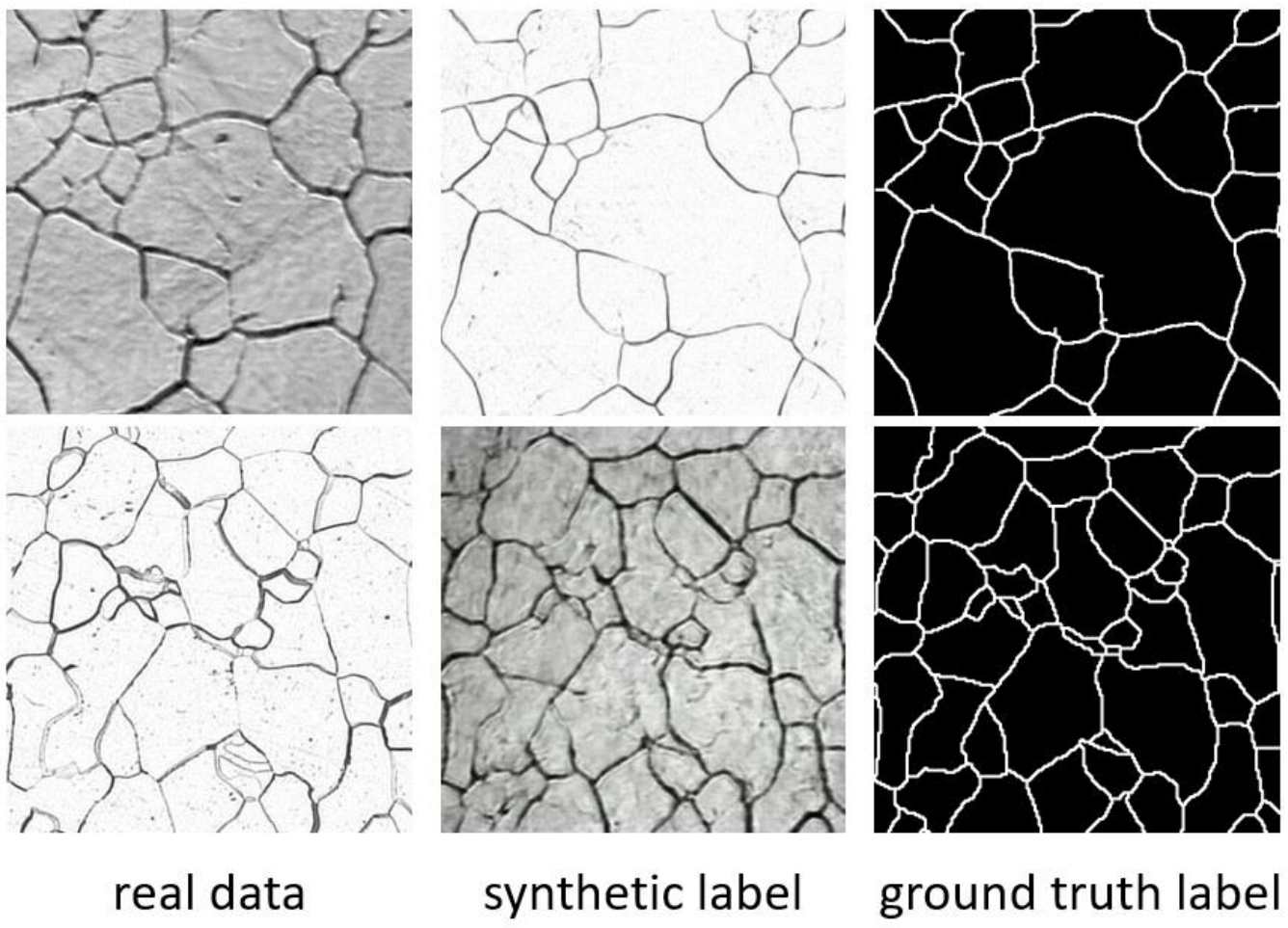

**Fig 5 The result of image style transfer.** The first row is austenite data, the second row is polycrystalline iron data.

In general, FedTransfer based on the characteristics of material microscopic image data, introduces transfer learning to share insensitive style information, reduce the influence of Non-IID data to the greatest extent under the premise of satisfying privacy constraints. Experiment shows that FedTransfer effectively improve the performance of the federation learning model on microscopic image segmentation task.

# RESULTS

### Datasets

We define a data point of every client as a pair of an image and its corresponding label. Throughout the paper, we consider the manually annotated labels as ground true (or true labels). The datasets we used consists of two parts: austenite dataset and polycrystalline iron dataset, which have different appearance in order to simulate the Non-IID of different clients' local data.

***Austenite dataset*.** The austenite dataset contains a total of 223 optical images of Austenite with a resolution of 1024 × 1024 pixels, of which the ground truth has 2 semantic classes (grain and grain boundary) and is manually labeled by material scientists. The original images were pre-processed into 892 small images with a size of 400 × 400 pixels and we used all the images for experiment. Finally, the training set consists of 560 images, the validation set consists of 140 images and the test set consists of 192 images.

***Polycrystalline iron dataset*.** The polycrystalline iron dataset contains a total of 136 serial section optical images of polycrystalline iron with a resolution of 2800 × 1600 pixels, of which the ground truth has 2 semantic classes (grain and grain boundary) and is manually labeled by material scientists. The original images were pre-processed into 2800 small images (patches) with a size of 400 × 400 pixels and we selected 882 images for experiment. Finally, the training set consists of 560 images, the validation set consists of 140 images and the test set consists of 192 images.

In our experiment, every client also used random erasing data augmentation methods in order to reduce the risk of over fitting.

## Federated learning setting

In our experiment, we set the amount of client participating in federated training as 2. In order to simulate the Non-IID of data across the clients, we deploy austenite dataset and polycrystalline iron dataset to 2 clients respectively where the first client only has austenite data while the other only has polycrystalline iron data. We set the maximum communication rounds of separate training as 40, as for federated training and centralized training which is set as 20. We choose the model which has the best performance on the average of every clients' test result as the final model.

---

**algorithm 1** FedTransfer

---

**Input:** $M$ is the set of clients participating in federated training;$K$ is the total communication round; $E$ is the number of local epochs; $B$ is the local minibatch size; $\eta$ is the learning rate; $n_i$ is the amount of local training data of client $i$; $n$ is the total amount of local training data of all clients.
**Output:** The trained model parameter $w^*$.

**function** FEDTRANSFER( )
    FEDERATEDIMAGESTYLETRANSFER( )
    $w^* \leftarrow$ FEDERATEDTRAINING( )
    **return** $w^*$
**end function**

**function** FEDERATEDIMAGESTYLETRANSFER( )
    **for** each client $i \in M$ **in parallel do**
        client $i$ use its data to train a pix2pix GAN $G_i$ which translate ground truth label to real image
        client $i$ upload $G_i$ to server
        server add $G_i$ to set $G$
    **end for**
    **for** each client $i \in M$ **in parallel do**
        server send $G$ to client $i$ except $G_i$
        client $i$ using $G$ and its ground truth label to generate synthetic data
        client $i$ put synthetic data into its local dataset
    **end for**
**end function**

**function** FEDERATEDTRAINING( )
    initalize $w^0$
    initalize $w^* = w^0$
    initalize $\ell_{min} = 9999$
    **for** each round $t$ = 1, 2, ..., $K$ **do**
        **for** each client $i \in M$ **in parallel do**
            ${w_i}^{t+1} \leftarrow$ CLIENTLOCALTRAINING$(k, w^t)$
            $w^{t+1} = \sum\limits_{i \in M} \frac{n_i}{n} {w_i}^{t+1}$
            $\ell_i \leftarrow$ client $i$ calculate loss of $w^t$ with its validation dataset
        **end for**
        $\ell_{mean} = \frac{1}{|M|} \sum\limits_{i \in M} \ell_i$
        **if** $\ell_{mean} < \ell_{min}$ **then**
            $\ell_{min} = \ell_{mean}$
            $w^* = w^t$
        **end if**
    **end for**
    **return** $w^*$
**end function**

**function** CLIENTLOCALTRAINING$(k, w)$
    $\beta \leftarrow$ (split local data into batches of size $B$)
    **for** each epoch $i$ = 1, 2, ... , $E$ **do**
        **for** batch $b \in \beta$ **do**
            $w \leftarrow w - \eta \bigtriangledown \ell(w; b)$
        **end for**
    **end for**
    **return** $w$ to server
**end function**

---

## Evaluation models and metrics

***Model setting***. There are two deep learning-based models in our experiment: image style transfer model and image segmentation model.

For image style transfer model, we used pix2pix2HD[17] to transform simulated label to synthetic image. During the training stage, we set the batch size equal to 2 with an initial learning rate of $2\times10^{-4}$ in 50 epochs.

For image segmentation model, we used U-net[20], the most commonly used supervised learning method in the field of materials and medical image processing[21]. During the training stage, we perform 5 folds cross validation to evaluate the model performance. We set the batch size equal to 8 and optimized the model by Adam with an initial learning rate of $1\times10^{-4}$.

The experiment were performed on a GeForce GTX 1080Ti GPU with 12 GB memory.

***Metrics***. It is important to choose effective metrics to evaluate the performance of algorithms we proposed in this task. In order to evaluate the recognition accuracy of the grain level and the segmentation accuracy of pixel clustering, we finally decided to use four effective metrics: mean average precision (MAP)[22]-[23] , mean variation of information(MVI)[24] and adjusted rand index (ARI)[25][27].

MAP is a classical measure in image segmentation and object detection tasks. In this paper, we evaluate it at different intersection over union (IoU) thresholds. The IoU of a proposed set of object pixels and a set of true object pixels is calculated as:

$$\mathrm{IoU}(A,B)=\frac{A\cap B}{A\cup B} \tag{1}$$

The metric sweeps over a range of IoU thresholds at each point, calculating an average precision value. The threshold values range from 0.5 to 0.95 with a step size of 0.05: (0.5, 0.55, 0.6, 0.65, 0.7,0.75, 0.8, 0.85, 0.9, 0.95). In other words, at a threshold of 0.5, a predicted object is considered a “hit” if its IoU with a ground truth object is >0.5. Generally, it can be considered that the segment is correct when its IoU is >0.5. The other higher value is aimed at ensuring the correct results.

At each threshold value t, a precision value is calculated based on the number of true positives (TP), false negatives (FN), and false positives (FP) resulting from comparing the predicted object to all ground truth objects. The average precision of a single image is then calculated as the mean of the above precision values at each IoU threshold.

$$\text{Average precision}=\frac{1}{|\text{threshold}|}\sum_{t}\frac{TP(t)}{TP(t)+FP(t)+FN(t)} \tag{2}$$

MVI is an information theoretic criterion for comparing two partitions, or clusterings, of the same data set. It measures the amount of information lost and gained in changing from clustering X to clustering Y[24]. variation of information is defined as:

$$VI(X,Y) = H(X|Y) + H(Y|X) \tag{3}$$

For image segmentation task, Y is the ground-truth segmentation, $H(Y|X)$ is the amount of under-segmentation of Y and $H(X|Y)$ is the amount of over-segmentation. In other words, a perfect over-segmentation will have $H(Y|X)=0$ and a perfect under-segmentation will have $H(X|Y)=0$.

**Table 1** Contingency table

| | $Y_1$ | $Y_2$ | ... | $Y_s$ | Sums |
|---|---|---|---|---|---|
| $X_1$ | $n_{11}$ | $n_{12}$ | ... | $n_{1s}$ | $a_1$ |
| $X_2$ | $n_{21}$ | $n_{22}$ | ... | $n_{2s}$ | $a_2$ |
| ... | ... | ... | ... | ... | ... |
| $X_r$ | $n_{r1}$ | $n_{r2}$ | ... | $n_{rs}$ | $a_r$ |
| Sums | $b_1$ | $b_2$ | ... | $b_s$ | |

ARI is the corrected-for-chance version of the rand index (RI), which is a measure of the similarity between two data clustering methods[25][27]. From a mathematical standpoint, ARI or RI is related to accuracy. In addition, image segmentation can be considered a clustering task that splits all pixels in an image into n partitions or segments. Given a set S of n elements (pixels) and two groupings or partitions of these elements, namely, $X = \{x_1, \ldots, x_r\}$ (a partition of S into r subsets) and $Y = \{y_1, \ldots, y_s\}$ (a partition of S into s subsets), the overlap between X and Y can be summarized in a contingency table [nij], where each entry nij denotes the number of objects in common between Xi and Yj: $n_{ij} = |X_i \cap Y_j|$. For the image segmentation task, X and Y can be treated as ground truth and predicted result, respectively.

The ARI is calculated as follow:

$$\text{ARI} = \frac{\sum_{ij}\binom{n_{ij}}{2} - \left[\sum_i\binom{a_i}{2}\sum_j\binom{b_j}{2}\right]/\binom{n}{2}}{\frac{1}{2}\left[\sum_i\binom{a_i}{2} + \sum_j\binom{b_j}{2}\right] - \left[\sum_i\binom{a_i}{2}\sum_j\binom{b_j}{2}\right]/\binom{n}{2}} \tag{4}$$

where $n_{ij}$, $a_i$, $b_j$ are values from the contingency table and $\binom{n}{2}$ is calculated as $n(n-1)/2$.

**Image segmentation by the proposed method in federated learning setting**

We first compared the performance of image segmentation algorithms in two cases: separate training and central training. Separate training means each client trains the personalized model only with its local training dataset without collaborating with other clients, while central training means directly training the model with the global training dataset which consist of the local training dataset of all clients. As shown in Table 2 (bold ones are the best method for corresponding metrics on different test datasets). Comparing with centralize training, separate training causes each client to perform better only on its local test dataset while perform poorer on the global test data set due to the lack of collaborations between clients. Besides, we notice that the performance of centralize training and separate training on local test dataset is very close, it indicates that centralize training leads to generalized model while separate training leads to specialized model. It is more important for us to train a generalized model, but directly using centralize training will cause some privacy issues.

After that, we explored the performance of the algorithm we proposed by comparing with other methods. Table 3 shows the performance of FedAvg[18] and FedTransfer. FedAvg is a baseline method which different client directly using FedAvg algorithm to perform collaboration while get raid of the privacy issue. FedTransfer is the method we proposed, there are two stages in our approach: image style transfer and federated training, both of which need to be trained. Result shows that FedTransfer perform better than FedAvg on all the metric on both global test dataset and local test dataset, besides Fig 6 shows the over all performance of separate, centralize and federated training, FedTransfer also shows better result than separate training on global test dataset as well as comparable result with separate training on both austenite and polycrystalline iron test dataset. It indicates that federated learning can improve the generalization ability of the segmentation model and our image style transfer method is useful to alleviate the performance degradation caused by the Non-IID of different clients' local dataset.

**Table 2** Performance of image segmentation in Separate training and centralize training cases

| Methods | Performance on austenite test set | | | Performance on polycrystalline iron test set | | | Average Performance on global test set | | |
|---|---|---|---|---|---|---|---|---|---|
| | MAP ↑ | MVI↓ | ARI↑ | MAP ↑ | MVI↓ | ARI↑ | MAP ↑ | MVI↓ | ARI↑ |
| Separate ( Austenite dataset) | **0.585 ± 0.011** | **0.190 ± 0.006** | **0.821 ± 0.005** | 0.476 ± 0.020 | 0.375 ± 0.034 | 0.702 ± 0.019 | 0.530 ± 0.013 | 0.283 ± 0.015 | 0.762 ±0.010 |
| Separate (Polycrystalline iron dataset) | 0.165 ± 0.067 | 1.636 ± 1.185 | 0.358 ± 0.147 | **0.613 ± 0.012** | **0.133 ± 0.008** | **0.807 ±0.010** | 0.389 ± 0.030 | 0.884 ± 0.589 | 0.582 ± 0.072 |
| Centralize | 0.580 ± 0.021 | 0.192 ± 0.011 | 0.817 ± 0.015 | 0.600 ± 0.008 | 0.144 ± 0.006 | 0.799 ± 0.007 | **0.590 ± 0.013** | **0.168 ± 0.008** | **0.808 ± 0.010** |

**Table 3** Performance of image segmentation in federated setting

| Methods | Performance on austenite test set | | | Performance on polycrystalline iron test set | | | Average Performance on global test set | | |
|---|---|---|---|---|---|---|---|---|---|
| | MAP ↑ | MVI↓ | ARI↑ | MAP ↑ | MVI↓ | ARI↑ | MAP ↑ | MVI↓ | ARI↑ |
| FedAvg | 0.480 ± 0.014 | 0.223 ± 0.006 | 0.747 ± 0.013 | 0.539 ± 0.014 | 0.250 ± 0.025 | 0.747 ± 0.014 | 0.510 ± 0.006 | 0.236 ± 0.013 | 0.747 ± 0.004 |
| FedTransfer | **0.559 ± 0.018** | **0.195 ± 0.004** | **0.809 ± 0.011** | **0.586 ± 0.010** | **0.162 ± 0.012** | **0.801 ± 0.007** | **0.575 ± 0.009** | **0.178 ± 0.007** | **0.807 ± 0.006** |

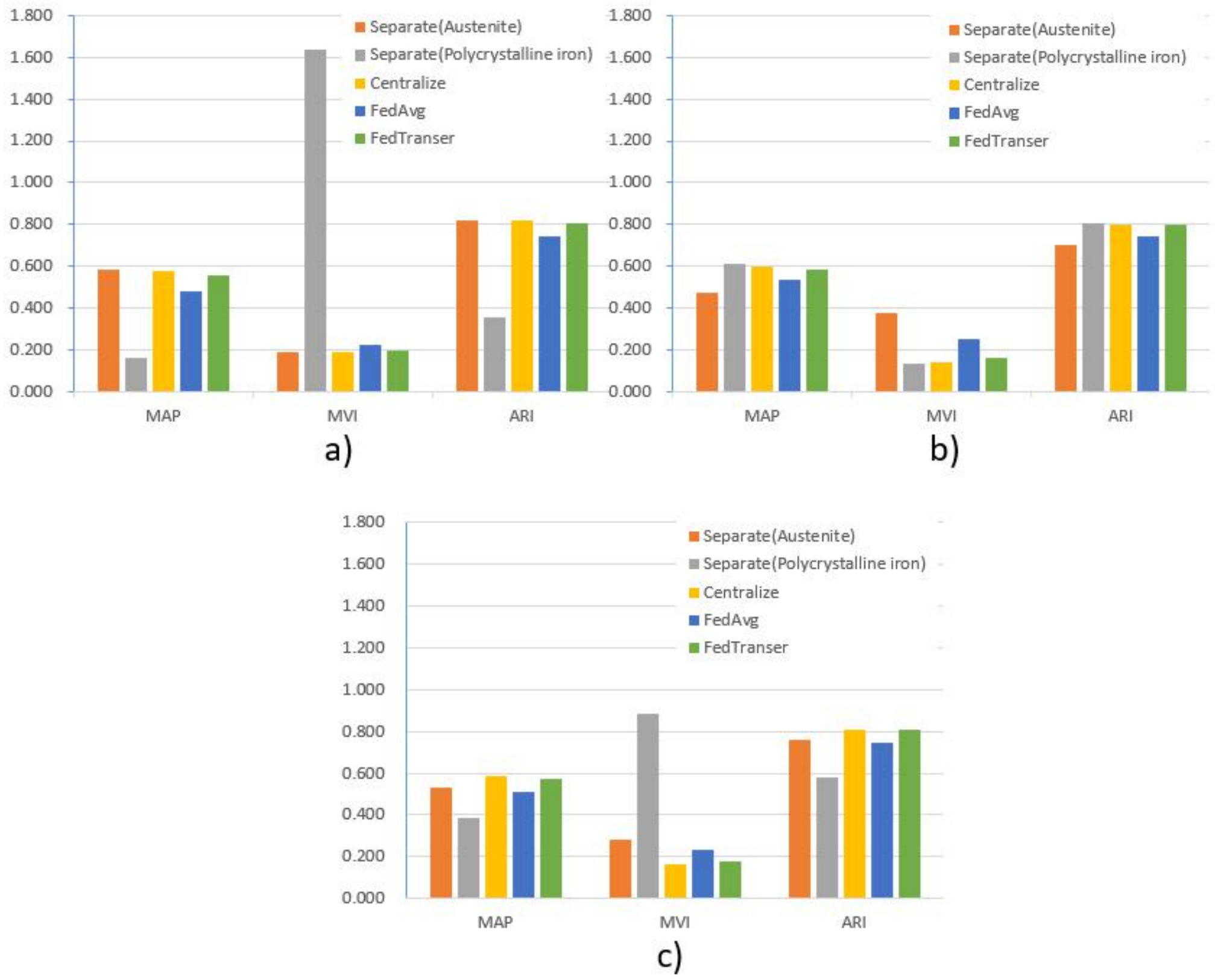

**Fig 6 Performance of separate, centralize and federated training.** a) shows the performance on austenite test set; b) shows the performance on polycrystalline iron test set; c) shows the average performance on global test set

# Discussion

Modern machine learning methods, such as deep learning, have outstanding prediction performance when the model is trained with a sufficiently large amount of data[13]. However, most applications in materials science troubled by isolated data island where each researcher can only use its own data to train models with limited generalization because of the data privacy issue.

In recent year, the federated learning framework provides a feasible solution for machine learning based on data privacy protection. However, in the field of materials science, different data preparation methods often bring Non-IID across different client participate in the federated training which brings worse performance and slower convergence.

In the present work, we propose a new material microscopic image federation learning scheme. By sharing the synthesized virtual image, the insensitive style information is shared without leaking the important structural information. Based on the characteristics of material microscopic image data, our methods reduce the influence of Non-IID data to the greatest extent under the premise of satisfying privacy.

This work has demonstrated the viability of the combination of simulated and real experimental data, suggesting that simulated data (after performing image style transfer from the real data) could be useful in data mining or machine learning problems. We believed that the proposed strategy could be easily applied to other material image processing task.

Our next step will on two issues: 1). Using the model quantification technique to compress the style transfer model to reduce the bandwidth consumption during the transmission of the style transfer model. 2)To investigate a more efficient transfer learning technique for this segmentation task which can fully extract the style information of material image data so as to further improve algorithm's performance.

## Methods

### Experimental

A commercial hot-rolled iron plate with a purity of >99.9 wt% was used in this work. The plate was forged into a round bar with a diameter of 30 mm and then fully recrystallized by annealing at 1153 K for 3 h. The samples for metallographic characterization were spark cut from the bar. One surface of the samples was mechanically polished for a fixed time, and metallographic photographs were then taken with an optical microscope after etching using a 4 vol% nital solution. The steps of the polish-etching photograph above were repeated to obtain serial section photographs. The average thickness interval between the two sections is ~1.8 μm.

### Federated Learning Setting

Recently, Federated learning has received significant interest from research and applied perspectives, it is a machine learning setting where many clients collaboratively train a model under the coordination of a central server without centralize their training data. It embodies the principles of focused collection and data minimization, and can mitigate many of the systemic privacy risks and costs resulting from traditional, centralized machine learning[28]. According to the attributes of the clients participating in the federated learning, federated learning can be divided into two categories: Cross-silo federated learning and Cross-device federated learning.

***Cross-device federated learning.*** The clients are a very large number of mobile or IoT devices( up to $10^{10}$ clients) under the setting of cross-device federated learning. In this setting, communication cost is often the primary bottleneck. During one communication round, only a fraction of clients are available and the clients are stateless as well as highly unreliable[28].

***Cross-silo federated learning.*** The clients are different organizations or geo-distributed data centers (typically 2 - 100 clients) under the setting of cross-silo federated learning. In this setting, computation or communication cost is often the primary bottleneck. During one communication round, all the clients are available and the clients are stateful and reliable[28].

In materials informatics filed, every researcher who owns the training data can be seen as a client, the situation is more in line with the cross-silo federated learning setting.

### Image style transfer

We used the image style transfer model to perform federated style transfer by extracting the style information from real images. Specifically, we choose the conditional GAN (pix2pixHD)[17][29] to carry out the image-to-image transformation. A GAN consists of a generator G and a discriminator D. Assume that x is the source image and y is the target image. The generator G produces an output image G(x) while the discriminator D distinguishes G(x) from the corresponding y. Both modules are optimized by adversarial training to make the "fake" G(x) closer to the "true" y.

The objective function of a conditional GAN can be expressed as:

$$L_{cGAN}(G,D) = E_{x,y}[logD(x,y)] + E_x[log(1 - D(x,G(x)))] \tag{5}$$

where G tries to minimize this objective and D tries to maximize it.

$$G^* = \arg\min_G \max_D L_{cGAN}(G,D) \tag{6}$$

The final objective is as follows:

$$G^* = \arg\min_G \max_D L_{cGAN}(G,D) + E_{x,y}[||y - G(x)||_1] \tag{7}$$

The second term is the L1 loss which restrict the generator output to the ground truth in an L1 sense[30].

We performed image-to-image transfer to perform federated style transfer, as shown in Fig 5. clients' generator used the encoding–decoding network U-net[20] while the discriminator calculates the loss of local patches between the output and ground truth to represent the consistency of local details[31]. In the training stage, client use an image of manually annotated ground truth label as input and the generator is trained to recover the real image and the discriminator is trained to distinguish the "real ones" versus the generated "synthetic ones". During the training, the generator acquires the ability to extract the style information of a given image. After receiving other clients' generative model, each client perform inference on the images of its own label to generate synthetic image data.

# Acknowledgements

This work was supported by the National Natural Science Foundation of China under Grant 6210020684 and Grant 61873299, Scientific and Technological Innovation Foundation of Shunde Graduate School of USTB under Grant BK19AE034 and Grant BK20AF001 and Grant BK21BF002, and Fundamental Research Funds for the Central Universities of China under Grant 00007467 and Grant FRF-TP-19-043A2. The computing work is supported by USTB MatCom of Beijing Advanced Innovation Center for Materials Genome Engineering.